# Use of Set Shaping theory in the development of locally testable codes

Solomon Kozlov[1]

**Abstract:** in developing locally testable codes, information is added to the coded message by creating redundancy in the codewords. In this article, we propose an alternative method in which redundancy is introduced on the message that must be transmitted before its encoding and not in the codewords. This approach exploits the Set Shaping Theory whose goal is the study the bijection functions $f(X^N) = Y^{N+K}$ that transform a set of strings into a set of equal size made up of strings of greater length. In this way, this type of function transforms the independent variable $x$ into the dependent variable $y$ whose emission probability is conditioned by the previously emitted variables. Thus, if the decoder decodes a symbol associated with a conditional probability equal to zero, we detect an error in the message. If the function $f$ used is the one that minimizes the average information content, we develop a code that can be tested efficiently. In fact, it is observed, in terms of compression, that the greater length of the strings is compensated by the fact of having chosen the strings with less entropy.

**Introduction**

The locally testable codes LTC [1] and [2] are error-correcting codes that allow to test some bits and reject the transmitted words with a probability proportional to their distance from the code (level of redundancy).

The Set Shaping Theory studies the bijection functions $f: X^N \to Y^{N+K}$ with $K$ and $N \in \mathbb{N}^+$, $|X^N| = |Y^{N+k}|$ and $Y^{N+K} \subset X^{N+K}$, which transform a set of strings $X^N$ into a set $Y^{N+K}$ made up of strings of greater length. Consequently, the probability of generating the sequences belonging to the set $X^{N+K}$ that do not belong to $Y^{N+K}$ is zero. In this way, this type of function transforms the independent variable $x$ into the dependent variable $y$ whose emission probability is conditioned by the previously emitted variables $p(y_j \mid y_1 \cap y_2 \ldots \cap y_{j-1})$. Thus, there will be situations where $p(y_j \mid y_1 \cap y_2 \ldots \cap y_{j-1})$ will be equal to zero. In these cases, if the decoder decodes the $y_j$ symbol, we detect an error in the message.

If the function $f$ is chosen in such a way as to select from the set $X^{N+K}$ the strings with the lowest information content, we develop a code that can be tested efficiently in the points where $p(y_j \mid y_1 \cap y_2 \ldots \cap y_{j-1})$ is zero. In fact, it is observed, in terms of compression, that the greater length of the strings is compensated by the fact of having chosen from the set $X^{N+K}$ the strings with less entropy.

**Methods**

To understand the reported results it is essential to define some functions used in the Set Shaping Theory [3], which are based on the theory developed by C.E.Shannon [4].

Given a source defined by an ensemble $X = (x; A; P)$, where $x$ is the value of the random variable, $A = \{a_1, a_2, \ldots \ldots a_I\}$ are the possible values of $x$ (states) and $P = \{p_1, p_2 \ldots \ldots p_I\}$ is the probability distribution of the states $P(a_i) = p_i$ with $\sum_{i=1}^{I} pi = 1$.

The entropy of $X$, denoted $H$, is defined as:

[1] Author correspondence: solomon.kozlov@mailfence.com

$$H(X) = -\sum_i p_i \log_b p_i$$

We call $X^N$ the set that contains all possible strings $\boldsymbol{x} = \{x_1, \ldots, x_j, \ldots, x_N\}$ generated by $X$. The information content of the string $\boldsymbol{x}$ is defined as follows:

$$I(\boldsymbol{x_i}) = -\sum_{j=1}^{N} \log_2 p(x_j)$$

The probability $P(\boldsymbol{x_i})$ that the source $X$ generates the sequence $\boldsymbol{x_i}$ is:

$$P(\boldsymbol{x_i}) = \prod_{j=1}^{N} p(x_j)$$

**Definition 1:** *We call $f$ the bijection function on the set $X^N$ defined as*:

$$f: X^N \to Y^{N+k}$$

$K, N \in \mathbb{N}^+$, $|X^N| = |Y^{N+k}|$ and $Y^{N+K} \subset X^{N+K}$.

$f(\boldsymbol{x}) = \boldsymbol{y}$ with $\boldsymbol{x} = \{x_1, \ldots, x_N\}$ and $\boldsymbol{y} = \{y_1, \ldots, y_{N+K}\}$, $\boldsymbol{x} \in X^N$ and $\boldsymbol{y} \in Y^{N+K}$

**Remark 1:** The function $f$ defines from the set $X^{N+K}$ a subset of size equal to $|X|^N$. This operation is called "Shaping of the source" and the parameter $K$ is called the shaping order of the source.

Since the set $Y^{N+K}$ is a subset of the set $X^{N+K}$, the probability $P(y)$ of generating a string $\boldsymbol{y} \in X^{N+K} \wedge \boldsymbol{y} \notin Y^{N+K}$ is zero. Consequently, the function $f$ transforms the independent random variable $x$ into the dependent variable $y$ whose emission probability is conditioned by the previously emitted variables $p(y_j \mid y_1 \cap y_2 \ldots \cap y_{j-1})$. In this way, the emission probability $P(y)$ of the string $y$ with $f(\boldsymbol{x}) = \boldsymbol{y}$ is:

$$P(\boldsymbol{y}) = \prod_{j=1}^{N} p(x_j)$$

$$\prod_{j=1}^{N} p(x_j) = p(y_1) p(y_2 \mid y_1) \ldots p(y_{N+K} \mid y_1 \cap y_2 \ldots \cap y_{N+K-1}) \quad (1)$$

When $\boldsymbol{y} \in X^{N+K} \wedge \boldsymbol{y} \notin Y^{N+K}$ we have $P(y)=0$ consequently, there will be at least one $p(y_j \mid y_1 \cap y_2 \ldots \cap y_{j-1}) = 0$. Therefore, if during the decoding we obtain the symbol $y_j$ associated with a $p(y_j \mid y_1 \cap y_2 \ldots \cap y_{j-1}) = 0$, we are sure of the presence of an error in the data transmission.

**Remark 2:** in this way, we try to make the code locally testable before its encoding takes place. This point is fundamental, because we use the function $f$ which transforms the set $X^N$ into the set $Y^{N+K}$ composed of $|X^N|$ strings with less information content belonging to $X^{N+K}$. Therefore, by performing the entropy coding after applying the function $f$, we can exploit the characteristic of the set $Y^{N+K}$ of being made up of the strings with the lowest entropy belonging to $X^{N+K}$.

**Definition 2:** *We call $f_m$ the bijection function on the set $X^N$ defined as*:

$$f_m: X^N \to Y^{N+k}$$

With $K, N \in \mathbb{N}^+$, $|X^N| = |Y^{N+k}|$, $Y^{N+K} \subset X^{N+K}$, $X^{N+K} - Y^{N+K} = C^{N+K}$, $\forall \boldsymbol{y} \in Y^{N+K}$ and $\forall \boldsymbol{c} \in C^{N+K}$ $I(\boldsymbol{y}) < I(\boldsymbol{c})$ and $I(y_i) < I(y_{i+1}) \forall \boldsymbol{y} \in Y^{N+K}$.

The function $f_m$ transforms the set $X^N$ into the set $Y^{N+k}$ composed of $|X^N|$ strings with less information content belonging to $X^{N+K}$. Consequently, each string belonging to the complementary set of $Y^{N+K}$ has a greater information content than any string belonging to $Y^{N+k}$.

**Definition 3:** *we call the average information content of a sequence generated by a source $X = (x; A; P)$ the summation of the product between the information content of the sequences belonging to $X^N$ is their probability*:

$$I(\boldsymbol{x}) = \sum_{i=1}^{|X|^N} P(\boldsymbol{x_i}) I(\boldsymbol{x_i}) \quad (2)$$

Applied the function *f*, the problem arises of how to calculate the new entropy of the transformed source *f(X)=Y*. In fact, the new source *Y* is very different from source *X* since it consists of dependent variables *y* associated with the conditional probabilities $p(y_j \mid y_1 \cap y_2 \ldots \cap y_{j-1})$. This problem is solved by calculating the average information content (2). In fact, the Asymptotic Equipartition Principle [5] tells us that when the length of the string increases the contribution to the value of the function (2) derives almost exclusively from the strings belonging to the typical set [6]. With typical set we mean the set of strings whose information content is close to *NH(X)*. Consequently, when *N* tends to infinity *I(x)* tends to *NH(X)*.

**Results**

To understand a possible use of the $f_m$ function, in the development of a locally testable code, it is necessary to study how the average information content varies with respect to the parameters *|A|* and *K*.

Of particular interest is the shaping order of the source *K*, in fact this parameter represents the difference in length between the sequences belonging to $Y^{N+K}$ and the sequences belonging to $X^N$. Hence, the parameter *K* also defines the minimum Hamming distance between a string $y \in Y^{N+K} \wedge y \notin X^{N+K}$ and a string $x \in X^{N+K} \wedge x \notin Y^{N+K}$. The Hamming distance between two strings represents the number of positions where the corresponding symbols are different.

This characteristic of parameter *K* can be used to estimate the probability with which a single variation of a symbol in the transmitted sequence leads to a string $x \in X^{N+K} \wedge x \notin Y^{N+K}$. When *N* is large and the function $f_m$ is applied, this type of probability can be estimated from the following ratio:

$$\frac{K}{N+K} \qquad (3)$$

Now, we begin by analyzing how the average information content (2) varies as a function of *|A|* and *K*. In this analysis, we consider a source defined by an ensemble $X = (x; A; P)$ which generates strings of length 100 with *|A|* varying between 2 and 5 and uniform probability distribution. On the strings generated by *X*, we will apply the function $f_m$ with *K={1,2}*. The exact calculation of *I(x)* and *I(y)* is very complex for messages with this length, so we estimate these values using the Mote Carlo method [7] and [8]. The data reported in Table 1 concern the simulation of $10^6$ strings of length 100 generated by source *X*.

In table 1, the first column shows the cardinality of *A*, the second column shows the value of *K*, the third column shows the value of *I(x)*, the fourth column shows the value of *I(y)* and finally the fifth column shows the difference *I(x)-I(y)*.

| \|A\| | K | I(x)  N = 100 | I(y)  N = 100 + K | I(x)-I(y) |
|---|---|---|---|---|
| 2 | 1 | 99.275 | 99.660 | -0.385 |
| 2 | 2 | 99.275 | 99.917 | -0.642 |
| 3 | 1 | 157.044 | 157.034 | 0.011 |
| 3 | 2 | 157.044 | 157.008 | 0.037 |
| 4 | 1 | 197.816 | 197.331 | 0.485 |
| 4 | 2 | 197.816 | 197.069 | 0.747 |
| 5 | 1 | 229.279 | 228.315 | 0.964 |
| 5 | 2 | 229.279 | 227.819 | 1.459 |

*Table 1: The average information content I(x) and I(y) in bits calculated for N = 100 and K = {1,2}.*

The data in table 1 show that for values of *|A| > 2* the average information content *I(y)* is less than *I(x)*. Of particular interest is the reduction of the value of *I(y)* when passing from *K=1* to *K=2* for *|A| > 2*. On the other hand, for *|A| = 2|, I(y)* increases, so increasing *K* does not solve the problem of having a value of *I(y)* greater than *I(x)*.

**Remark 3:** from table 1 it is also noted how the average information content approximates *NH(X)*. In fact, with *N=100* we have *I(x)=99.275* a value very close to *NH(X)=100*. Consequently, the contribution to formula (2) depends almost exclusively on strings with information content close to *NH(X)*. This data is fundamental because formula (2) is used to indirectly study the entropy of the transformed source.

Now, we study how the average information content varies by varying *K* and leaving the alphabet and the length of the string unchanged. Hence, we simulate a source $X = (x; A; P)$ with $|A| = 3$ generating strings of length 100 with uniform probability distribution. On the strings generated by *X*, we will apply the function $f_m$ with *K* variable from 1 to 5. In this case, in order to have an acceptable error, we need to increase the number of strings simulated by the Monte Carlo method to $10^7$.

In table 2, the first column shows the value of *K*, the second column shows the value of *I(x)*, the third column shows the value of *I(y)* and finally, the fourth column shows the difference *I(x)-I(y)*.

| K | I(x) N = 100 | I(y) N = 100 + K | I(x)-I(y) |
|---|---|---|---|
| 1 | 157.044 | 157.033 | 0.011 |
| 2 | 157.044 | 157.019 | 0.025 |
| 3 | 157.044 | 157.009 | 0.035 |
| 4 | 157.044 | 157.000 | 0.045 |
| 5 | 157.044 | 156.976 | 0.068 |

*Table 2: The average information content I(x) and I(y) in bits calculated for N = 100, |A| = 3 and K variable from 1 to 5.*

Analyzing the data in table 2, we note that if we increase *K* decreases *I(y)* and consequently increases the difference *I(x)-I(y)*. This result is very interesting because increasing *K* increases the probability of detecting an error (3). Obviously, this trend will reverse at some point, and increasing the parameter *K* will imply an increase in *I(y)*. To Understand for what value of *K*, we will have this change is not easy, because when we increase *K*, in order to have an acceptable error, we must also increase the number of simulated strings. Therefore, we soon arrive in a situation in which we have unacceptable simulation times. So far, we have always considered a source with a uniform probability distribution. Now, we try to understand what happens when the probability distribution is non-uniform. To do this analysis we compare the single values of $I(x_i)$ and $I(y_i)$ with $f_m(x_i) = y_i$. Consequently, having to calculate all the values of $I(x_i)$ and $I(y_i)$, we must consider strings of short length. For this reason, we will analyze the set $X^{10}$ consisting of strings of length 10 and $|A| = 3$, on which we will apply the function $f_m$ with $K = \{1,2,3\}$. The set $X^{10}$, thus defined, contains $3^{10} = 59049$ strings, a number of strings acceptable from the point of view of computation times and at the same time a number large enough to make negligible the contribution of strings with a content of much less information than *NH(X)*.

In figure 2, the solid line shows the values $I(x_i)$, the dashed line the values of $I(y_i)$ with *K=1*, the dashed line with dots the values of $I(y_i)$ with *K=2*, and the line with the dots the values of $I(y_i)$ with *K=3* in bits. The strings were sorted according to their information content in ascending order. The data in figure 1 show how the values of: $I(x_i)$, $I(y_i)$ $K = 1$, $I(y_i)$ $K = 2$ and $I(y_i)$ $K = 3$ oscillate between them. From the reported results, it can be seen how the information content $I(y_i)$ turns out to be less than $I(x_i)$ in points that change according to the value of *K* used in the transform. This result is very interesting because, on the basis of the probability distribution of the source to be encoded, we can choose the parameter *K* in which the typical set falls in the area where *I(y)<I(x)*. Obviously, this result needs many more studies; however this data shows us a possible solution to implement this method efficiently.

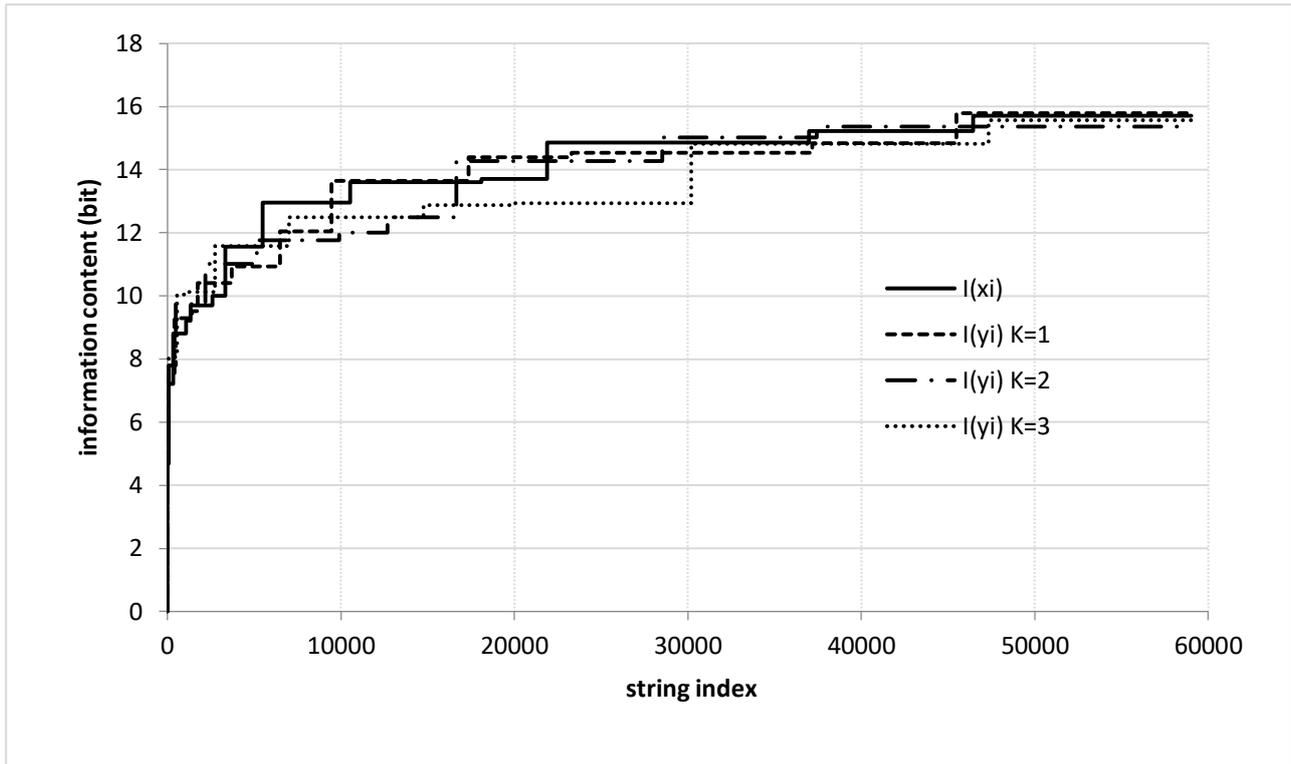

*Figure 1: comparison between $I(x_i)$ and $I(y_i)$ for N = 100, $|A| = 3$ and K = {1,2,3}.*

## Conclusion

In developing locally testable codes, information is added to the coded message by creating redundancy in the codewords. In this way, with a probability dependent on the increase in the length of the message, it is possible to detect the presence of errors in data transmission.

In this article, we have proposed an alternative method in which redundancy is introduced on the message that must be transmitted before its encoding and not in the codewords. This approach exploits the Set Shaping Theory whose goal is the study the bijection functions that transform a set of strings $X^N$ into a set $Y^{N+K}$ of equal size made up of strings of greater length. Consequently, through this function, the independent variable *x* is transformed into the dependent variable *y* with *y=f(x)*. Therefore, the conditional probability $p(y_j | y_1 \cap y_2 \ldots \cap y_{j-1})$ will be associated with the symbol $y_j$. Since $Y^{N+K}$ is a subset of $X^{N+K}$ there will exist values of $y_j$ with a conditional probability equal to zero. If in these cases, during the decoding phase, we detect the $y_j$ symbol, we are sure of the presence of an error in the transmitted message.

If the function $f(X^N) = Y^{N+K}$ is chosen so that the set $Y^{N+K}$ contains the strings with the least information content belonging to the set $X^{N+K}$ we are able, for some probability distributions that depend on the parameter *K*, to compensate the increase in the length of the string with a greater compression of the message. Consequently, we can exploit these results to increase the efficiency in the development of locally testable code.